\documentclass[9pt,conference]{IEEEtran}

\usepackage{algorithm}
\usepackage{algpseudocode}
\usepackage[table]{xcolor}

\usepackage[preprint]{waspaa25}

\usepackage{bm} 
\definecolor{myblue}{HTML}{DAE8FC}
\colorlet{lightblue}{myblue!75!white}

\title{SynSonic: Augmenting Sound Event Detection through Text-to-Audio Diffusion ControlNet and Effective Sample Filtering}

\name{
Jiarui Hai, 
Mounya Elhilali
}
\address{Department of Electrical and Computer Engineering\\ Johns Hopkins University, Maryland, USA}

\begin{document}

\maketitle

\begin{abstract}
Data synthesis and augmentation are essential for Sound Event Detection (SED) due to the scarcity of temporally labeled data. While augmentation methods like SpecAugment and Mix-up can enhance model performance, they remain constrained by the diversity of existing samples. Recent generative models offer new opportunities, yet their direct application to SED is challenging due to the lack of precise temporal annotations and the risk of introducing noise through unreliable filtering. To address these challenges and enable generative-based augmentation for SED, we propose SynSonic, a data augmentation method tailored for this task. SynSonic leverages text-to-audio diffusion models guided by an energy-envelope ControlNet to generate temporally coherent sound events. A joint score filtering strategy with dual classifiers ensures sample quality, and we explore its practical integration into training pipelines. Experimental results show that SynSonic improves Polyphonic Sound Detection Scores (PSDS1 and PSDS2), enhancing both temporal localization and sound class discrimination.
\end{abstract}

\section{Introduction}
Sound event detection (SED) is a technique to identify, classify, and temporally locate specific acoustic events within audio recordings \cite{mesaros2021sound}. It involves analyzing audio signals to detect occurrences of target sounds, such as speech, music, animal calls, or environmental sounds, along with their precise start and end times.

Recent advances in deep neural networks have led to significant progress in SED \cite{mesaros2021sound, nam2022frequency, miyazaki2020conformer}. However, training SED models typically requires strongly labeled data, which provide precise annotations of the exact start and end times of each sound event. Acquiring such datasets is challenging and costly due to the labor-intensive annotation process. To address this limitation, synthetic soundscapes with precise annotations have been proposed as alternative training resources for SED systems~\cite{serizel2020sound, turpault2019sound}. However, synthetic data introduces its own challenges. Although it enables the creation of numerous audio mixtures, the diversity of these mixtures is constrained by the range of available foreground and background samples. Furthermore, collecting high-quality, isolated foreground events remains a difficult task. When synthetic datasets are generated from limited sound samples, the resulting lack of diversity can lead to overfitting and impair the generalization capability of deep neural networks \cite{mesaros2021sound}. 

To mitigate these issues, data augmentation has emerged as a particularly effective solution. By artificially increasing the variability of training data, augmentation techniques can significantly enhance model robustness and improve generalization. Traditional data augmentation methods~\cite{lin23_interspeech, park2019specaugment, zhang2017mixup, xie2020unsupervised} have demonstrated strong performance in diversifying training signals without the need for additional manual annotations. However, these methods typically operate on existing data and are inherently limited by the diversity of the original samples, imposing an upper bound on the achievable variability.

Recently, generative models have been increasingly adopted for data augmentation in computer vision~\cite{wu2024image, li2024semantic, trabucco2023effective}, thanks to their ability to automatically produce diverse and realistic data. In the audio domain, text-to-audio (T2A) models have shown promising capabilities in generating sound clips containing single sound events, even when primarily trained on audio mixtures, and have been successfully used for data augmentation in sound extraction and classification ~\cite{wang2025soloaudio, ghosh2024synthio}. However, SED relies on precise temporal annotations, specifically, exact onset and offset times, to prepare strongly labeled data, which makes it challenging to directly apply the same augmentation strategies as sound separation and classification. Moreover, data selection is crucial when using generative models for data augmentation. Despite their ability to produce authentic data, generative models can still make mistakes, and indiscriminately incorporating generated samples may introduce noise and degrade performance. While models like CLAP \cite{wu2023large} can assist with filtering, relying on a single scoring model may introduce selection bias, potentially limiting the quality and diversity of generated samples and impairing SED’s robustness.

To bridge this gap and bring the benefits of generative model-based augmentation to SED, we introduce SynSonic\footnote{Source Code: 
 \url{https://github.com/JHU-LCAP/SynSonic}}, which combines controllable foreground sound generation with an effective and reliable sample selection method. The main contributions of this work are summarized as follows:

\textbf{(1) Controllable sound sample generation.} To generate tightly segmented sound events with precise temporal structures, we integrate an energy envelope-based control signal into a text-to-audio diffusion model~\cite{hai2024ezaudio} using ControlNet \cite{zhang2023adding}. This setup guides the generation process by enforcing target temporal patterns, enabling precise control over event timing during mixture synthesis.

\textbf{(2) Robust sample selection.} To automatically identify high-quality samples, we propose a dual-classifier joint ranking strategy that combines CLAP scores with classification logits from an AudioSet-based \cite{gemmeke2017audio} audio classification model \cite{dinkel2024scaling}. This method improves selection robustness, enhances sample quality, and reduces biases associated with single-score filtering.

\textbf{(3) Balanced Real/Synthetic fusion.}We conduct extensive experiments to establish practical guidelines for designing SynSonic, including analyses of foreground generation methods, data filtering settings, and mixture synthesis strategies. Additionally, we provide a usage recipe for effectively leveraging SynSonic, detailing configurations for Real/Synthetic data composition during training.

As a result, SynSonic consistently improves SED performance in both temporal localization and class discrimination metrics.




\begin{figure*}[t]
  \centering
  \includegraphics[width=\textwidth]{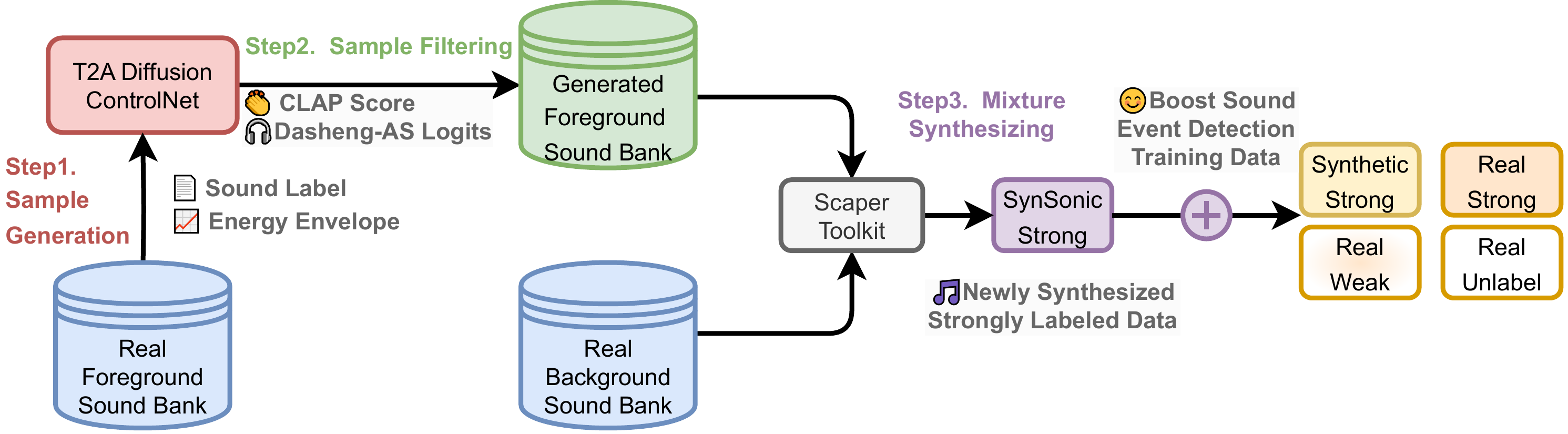}
  \caption{The overall framework of SynSonic.}
  \label{fig:framework}
\end{figure*}


\section{Method}
\label{sec:method}
In this section, we present the proposed data augmentation strategy, illustrated in \cref{fig:framework}, which comprises three key steps. First, foreground sound samples are generated using Energy-Envelope ControlNet to ensure both diversity and alignment with desired temporal characteristics. Second, the generated samples undergo filtering via a dual-classifier system to retain only high-quality and relevant examples. Finally, the selected samples are used to synthesize strongly labeled sound mixtures, thereby enriching the dataset and improving the robustness of SED model training.

\subsection{Soundbank Generation with Diffusion ControlNet}

\begin{figure}[t]
  \centering
  \includegraphics[width=\linewidth]{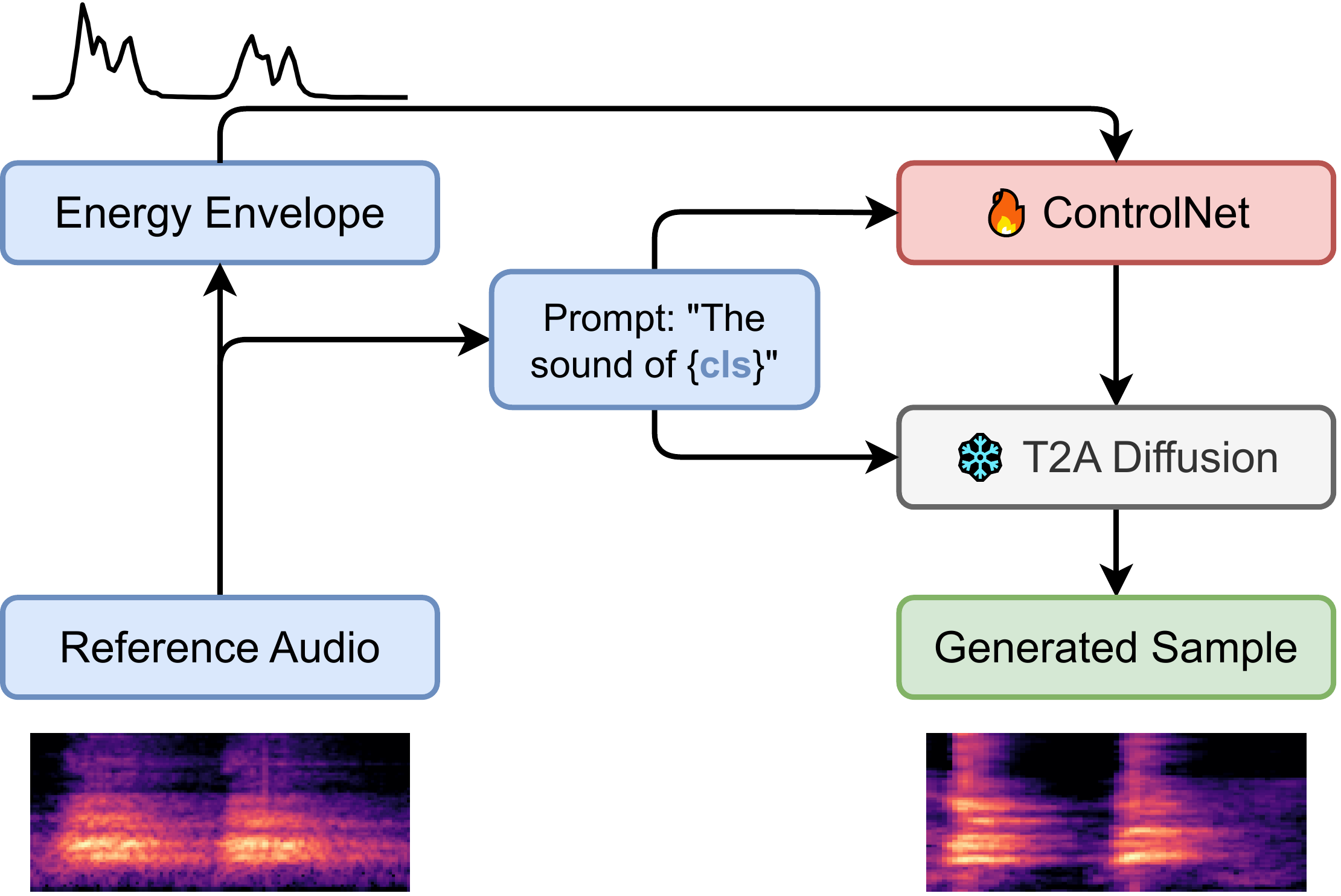}
  \caption{Illustration of the energy-envelope ControlNet generating new foreground samples, using the sound class \textit{dog} as an example.}
  \label{fig:controlnet}
\end{figure}

To address the challenge that foreground samples generated directly by T2A models often lack precise control over temporal attributes—such as timing and duration—which are critical for synthesizing strongly labeled data, we integrate a ControlNet-based mechanism into the generative process. ControlNet provides a structured approach that enables the model to be guided by external control signals in addition to the text prompt.

Specifically, since only the temporal structure needs to be controlled when generating foreground sound for SED, we adopt the energy envelope as the guiding signal, inspired by \cite{wu2024music, garcia2025sketch2sound}. This simple yet effective representation captures the temporal dynamics of sound events, including their onset, duration, and intensity progression. As a result, the diffusion model can generate audio that not only aligns with the semantic content of the text but also faithfully follows the intended temporal pattern. Importantly, it results in a well-defined onset-offset boundary.


We build the diffusion-based ControlNet on top of EzAudio \cite{hai2024ezaudio}, a recently released  T2A model based on diffusion transformers, which delivers superior audio quality and prompt coherence. Our implementation adopts a standard ControlNet architecture, in which the first half of the model’s blocks are duplicated. These ControlNet blocks are connected to the latter half of the main transformer through long skip connections. The energy-envelope is processed through a zero-initialized one-dimensional convolutional layer and subsequently added to the ControlNet input to provide temporal guidance.

Leveraging the generative capabilities of the diffusion ControlNet, as illustrated in \cref{fig:controlnet}, we observe that conditioning on energy references still yields diverse audio outputs. Although these outputs share similar temporal envelopes with the references, they exhibit substantial variation in frequency characteristics, underscoring the model’s ability to generate diverse and non-redundant samples from the reference samples. These generated samples, diverse in their frequency characteristics, can thus enhance data diversity and help improve the robustness of the SED model.

\subsection{Effective Sample Filtering with Dual Classifiers}

\begin{figure}[t]
  \centering
  \includegraphics[width=\linewidth]{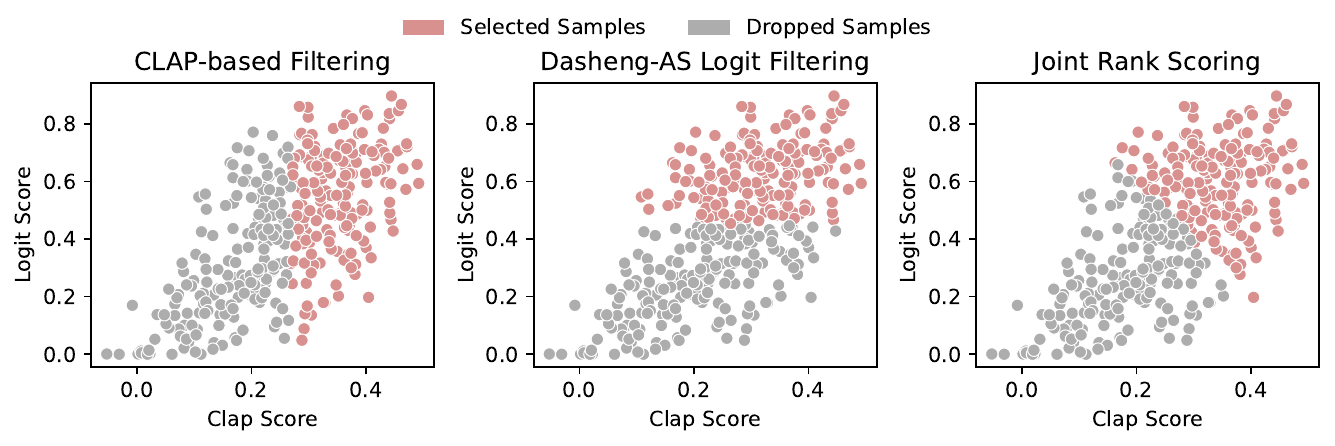}
\caption{Illustration of the filtering methods applied to samples generated for the sound class \textit{vacuum cleaner}.}
  \label{fig:filter}
\end{figure}

\begin{algorithm}[t]
\caption{Dual-Classifier Sample Filtering}
\label{alg:filtering}
\begin{algorithmic}[1]
\Require Generated samples $S$, CLAP scores $\{c_i\}$, classification logits $\{l_i\}$, weight $w$, top percentage $p\%$
\State Rank samples by CLAP score $\rightarrow$ R$_{\text{CLAP}}$
\State Rank samples by classification logit $\rightarrow$ R$_{\text{Cls}}$
\For{each sample $i$}
    \State Compute weighted rank score: $s_i = w \cdot \text{R}_{\text{CLAP},i} + (1 - w) \cdot \text{R}_{\text{Cls},i}$
\EndFor
\State Sort samples by $s_i$ and select top $p\%$
\State \Return Filtered samples
\end{algorithmic}
\end{algorithm}

Although the T2A model with ControlNet generally produces realistic sound samples, it can occasionally generate incorrect sound classes or unnatural audio. To ensure the quality of samples used for synthesizing strongly labeled data and to avoid introducing errors during SED model training, we adopt a filtering strategy to select only high-quality outputs. A CLAP-based model \cite{wang2025soloaudio, kong2024improving, wu2023large} can be employed to assess the similarity between the text prompt and the generated sample, helping to filter out unsuitable synthetic examples. However, this approach may introduce biases inherent to a single CLAP model.

To address this issue, we incorporate an additional audio classification model, Dasheng-AS \cite{dinkel2024scaling}, an AudioSet-based \cite{gemmeke2017audio} audio classification model, to provide a complementary score based on the predicted class logits. Since our focus is on foreground generation, and each foreground sample corresponds to a single class, this approach is well-suited. For each generated sample, both the CLAP similarity score and the classification logit are computed. As shown in \cref{alg:filtering}, to account for scale differences between these two metrics, samples within each class are independently ranked by each score. A weighted rank-based score is then calculated to determine the final ranking. Finally, only the top-ranked samples, selected according to a predefined percentage threshold, are retained for subsequent mixure creation.

As illustrated in \cref{fig:filter}, although the CLAP scores and Dasheng-AS logits exhibit an overall linear correlation, notable discrepancies between the two metrics remain. Some samples with high CLAP scores do not correspondingly achieve high logits, and vice versa, suggesting that relying on a single metric may introduce selection biases aligned with only one model’s perspective and lead to suboptimal sample quality. In contrast, the proposed joint ranking approach integrates both scores, mitigates such metric-specific biases, and ensures that only samples demonstrating consistently strong performance in both semantic similarity and classification confidence are selected.

\subsection{Mixture Synthesis}
After obtaining the generated and filtered foreground sound samples, we use the Scaper toolkit \cite{salamon2017scaper}, a soundscape synthesis library designed for creating labeled audio mixtures from foreground and background elements, to generate strongly labeled synthetic data. By default, we exclusively use the generated foreground sounds without mixing them with real foreground samples, while background and non-target sounds are sourced from real recordings \cite{fonseca2017freesound, turpault2019sound, mesaros2018multi}. Alternative settings, such as incorporating synthetic background sounds, are further explored in \cref{sec:mixing}.

This process involves two stages of synthesis, foreground generation via the diffusion transformer and mixture creation via Scaper, which differs from the synthetic strong subset used in DCASE dataset \cite{turpault2019sound}, where real foreground sounds are employed. To reflect this distinction, we refer to the dataset generated by our method as the SynSonic Strong subset. 

\section{Experimental Setups}

\subsection{SED Dataset}
We use the Domestic Environment Sound Event Detection (DESED) dataset \cite{turpault2019sound} to evaluate the effectiveness of the proposed method. This dataset focuses on 10 domestic sound classes and provides a comprehensive collection of recordings for training and evaluation. The training set includes real weakly labeled data (1,578 clips), real strongly labeled data (3,470 clips) \cite{hershey2021benefit}, synthetic strongly labeled data (10,000 clips), and real unlabeled data (14,412 clips). Additionally, a development set comprising 1,168 real clips is available for model development and validation. The SynSonic strong subset is used to augment the training data; specifically, we incorporate 10,000 samples generated using the method described in~\cref{sec:method}.

\subsection{Implementation Details}
We build the energy-envelope ControlNet on top of \textbf{EzAudio-L}\footnote{\url{https://huggingface.co/OpenSound/EzAudio}} and train it using the AudioCaps dataset~\cite{kim2019audiocaps}. For the energy envelope, we compute waveform energy, ensuring that the window and hop sizes are aligned with those of the audio latent representation. The ControlNet is trained with a batch size of 16, a learning rate of $1 \times 10^{-5}$, and for 10 epochs. During inference, we use a classifier-free guidance (CFG) scale of 3.5 and perform sampling with 50 diffusion steps.

For sample filtering, we employ the official checkpoints of Laion CLAP\footnote{\url{https://huggingface.co/laion/clap-htsat-fused}} and Dasheng-AS\footnote{\url{https://github.com/XiaoMi/dasheng}}, the latter being based on Dasheng-base and fine-tuned on AudioSet. Parameter selections are detailed in~\cref{sec:filter}.

For SED experiments, we adopt the FDY-CRNN model \cite{nam2022frequency} and follow the official training protocols, including Mean Teacher \cite{tarvainen2017mean}, Mixup \cite{zhang2017mixup}, and SpecAugment \cite{park2019specaugment}. Most hyperparameters, such as audio sample rate, mel-spectrogram settings, optimizer, learning rate, and loss weighting, remain consistent with the official implementation\footnote{\url{https://github.com/frednam93/FDY-SED}}, except for the batch size, which is adjusted to accommodate the additional synthesized data. The default batch composition is described and discussed in \cref{sec:batch}.

\subsection{Evaluation Metrics}
For evaluation, we adopt the PSDS1 and PSDS2 metrics \cite{bilen2020framework, mesaros2016metrics} to assess model performance, where the former emphasizes temporal localization and the latter focuses on event class discrimination. Following the DCASE baseline\footnote{\url{https://github.com/DCASE-REPO/DESED_task/tree/master}}, we apply a fixed median filter with a window size of 7 to smooth the model predictions, without further tuning this parameter in our experiments. To ensure robust and reliable results, we report \textbf{the average PSDS scores across three trials} with different random seeds.

\section{Results and Discussion}

\subsection{Comparison of Foreground Generation Methods}
We compare the ControlNet-based foreground generator with two alternative strategies: (1) simple T2A diffusion, as used in data augmentation for target sound extraction \cite{wang2025soloaudio}, and (2) audio-to-audio (A2A) diffusion, a simple method that aims to achieve controllable generation by conditioning on audio inputs.
 In the simple T2A diffusion approach, the generation length is determined based on reference samples, and silence at the beginning and end of the generated audio is trimmed to achieve the desired duration. However, because the generation process is not explicitly guided by temporal cues, the resulting samples often exhibit loose temporal structure, with unsmooth onsets and offsets as well as unintended pauses. In the A2A diffusion approach, inspired by the Img2Img framework~\cite{meng2021sdedit}, noise is added to the reference samples at a specific diffusion step (we use $75\%$ of the total diffusion steps). This encourages the generation of audio that preserves the temporal structure of the reference, while partially retaining its frequency characteristics, as they are not fully disentangled — unlike in the energy-envelope ControlNet. For fair comparison, we apply the same filtering and mixture synthesis strategies to both alternative methods as used in the proposed approach. Furthermore, we provide a model trained without any additional data for reference.

As shown in \cref{tab:generation}, simple T2A diffusion yields a clear improvement in the PSDS2 score, which emphasizes classification accuracy over temporal localization. However, it does not significantly improve PSDS1, which focuses on timing precision. This reflects the limitation discussed above: T2A diffusion lacks strong control over the temporal structure of generated samples. In contrast, A2A generation achieves better improvement in PSDS1 by producing samples with the desired temporal structure. However, it offers limited gains in PSDS2, as it tends to preserve certain frequency characteristics from the original audio, which reduces the diversity of generated samples. Compared to the two baselines, temporal ControlNet improves both PSDS1 and PSDS2, demonstrating its ability to balance precise temporal control and sample diversity, thereby enhancing model performance in both class discrimination and temporal localization.

\begin{table}[t]
    \centering
    \caption{Comparison of different foreground augmentation methods.}
    \setlength{\tabcolsep}{18pt}  
    \label{tab:generation}
    \begin{tabular}{l | c c}
        \toprule
        \textbf{Foreground Generator} & \textbf{PSDS1$\uparrow$} & \textbf{PSDS2$\uparrow$} \\
        \midrule
        None (Baseline)       & 0.4417 & 0.6639 \\
        \midrule
        T2A Diffusion      & 0.4507 & \underline{0.6859} \\
        A2A Diffusion                   & \underline{0.4556} & 0.6799 \\
        \rowcolor{lightblue} 
        Energy ControlNet  & \textbf{0.4641} & \textbf{0.6868} \\
        \bottomrule
    \end{tabular}
\end{table}

\subsection{Comparison of Sample Filtering Settings}
\label{sec:filter}

To investigate the effectiveness of the proposed sample filtering strategy, we conduct experiments using different weightings between the CLAP score and the audio classification score, ranging from purely CLAP-based to purely classifier-based filtering. In addition, we examine the impact of the filtering ratio, varying from no filtering to extreme filtering, where only the top $25\%$ of samples are retained.

As shown in \cref{tab:filter_ablation}, relying on a single filtering model leads to suboptimal performance in both PSDS1 and PSDS2. Specifically, using only CLAP (1.0/0.0) or only the audio classifier (0.0/1.0) results in noticeable degradation. In contrast, combining the two models improves performance, with equal weighting (0.5/0.5) achieving the best results. This suggests that integrating semantic relevance and audio classification confidence offers complementary benefits, enabling the selection of higher-quality, less noisy samples and thereby improving the SED model.

Meanwhile, we investigate the impact of the filtering threshold. Omitting the threshold results in almost no improvement over the baseline without newly synthesized data, highlighting the importance of data filtering. Furthermore, a lower threshold (top $75\%$) introduces noisy samples, while a higher threshold (top $25\%$) reduces data diversity, both leading to inferior improvements. By contrast, a threshold of $50\%$ achieves the best performance by effectively balancing quality and diversity.

\begin{table}[t]
    \centering
    \caption{Ablation study of sample filtering using different CLAP and AS weighting schemes.}
    \label{tab:filter_ablation}
    \begin{tabular}{c c c | c c}
        \toprule
        \textbf{Filter Ratio} & \textbf{CLAP Weight} & \textbf{AS Weight} & \textbf{PSDS1$\uparrow$} & \textbf{PSDS2$\uparrow$} \\
        \midrule
        Top 50$\%$ & 1.0  & 0.0   & 0.4565 & 0.6750 \\
        Top 50$\%$ & 0.7 & 0.3 & \underline{0.4590} & 0.6748 \\
        \rowcolor{lightblue} 
        Top 50$\%$ & 0.5   & 0.5  & \textbf{0.4641} & \underline{0.6868} \\
        Top 50$\%$ &  0.3 & 0.7 & 0.4573 & \textbf{0.6874} \\
        Top 50$\%$ & 0.0   & 1.0   & 0.4574 & 0.6784 \\
        \midrule
        None & / & / & 0.4436 & 0.6690 \\
        Top 75$\%$ & 0.5   & 0.5 & 0.4500 & 0.6776 \\
        Top 25$\%$ & 0.5   & 0.5  & 0.4589 & 0.6802\\
        \bottomrule
    \end{tabular}
\end{table}

\subsection{Discussion}

In this section, we present additional configurations of the proposed augmentation strategy explored during method development.

\subsubsection{Soundbank Generation and Mixture Synthesis}
\label{sec:mixing}

As shown in \cref{tab:mixing}, we compare two alternative strategies: (1) including real foreground audio when creating new strongly labeled mixtures, and (2) adding generated background or non-target sounds using the same generation strategy as for foreground samples. Incorporating real foreground audio leads to performance degradation, as these foreground samples have already been used in the original synthetic strong subset. Reusing them in newly synthesized mixtures increases the risk of overfitting, ultimately impairing the model's generalization to unseen data during evaluation. Meanwhile, since the SED model focuses on detecting target foreground sounds and the real background soundbank is already large and diverse, adding generated background samples does not offer an additional performance benefit. 

\begin{table}[t]
    \centering
    \caption{Comparison of different foreground mixing strategies.}
    \label{tab:mixing}
    \setlength{\tabcolsep}{17pt}  
    \begin{tabular}{l | c c}
        \toprule
        \textbf{Mixing Strategy} & \textbf{PSDS1$\uparrow$} & \textbf{PSDS2$\uparrow$} \\
        \midrule
        \rowcolor{lightblue} 
        Generated Foreground Only       & \textbf{0.4641} & \textbf{0.6868}\\
        \quad+ Real Foreground               & 0.4484 & 0.6826 \\
        \quad+ Generated Background          & 0.4613 & 0.6866 \\
        \bottomrule
    \end{tabular}
\end{table}



\subsubsection{Batch Composition}
\label{sec:batch}

As we introduce new strongly labeled subset during training, we explore different batch composition strategies to integrate it effectively. To better accommodate the new samples and understand their impact, we first reduce the proportion of each existing subset and fill the freed space with generated data. As shown in Table~\ref{tab:batch}, incorporating generated samples significantly improves performance when replacing synthetic strongly labeled data, which tends to lack foreground diversity. In contrast, replacing real data, whether it is strongly labeled, weakly labeled, or unlabeled, yields similar improvement on PSDS2, but minor improvements in PSDS1, suggesting the essential role of real data and the necessity of retaining it during training. In the final configuration, we directly add the generated samples on top of the original batch without substantially reducing the real data proportion. This strategy, which slightly lowers the relative weight of other subsets, achieves the best PSDS1 performance by increasing the diversity of strongly labeled data while maintaining sufficient real data.

\begin{table}[t]
    \centering
    \caption{Ablation study on training batch composition. The composition ratio indicates the proportion of: SynSonic strong / synthetic strong / real strong / real weak / real unlabeled samples in each batch.}
    \label{tab:batch}
    \setlength{\tabcolsep}{6pt}  
    \begin{tabular}{l c | c c}
        \toprule
        \textbf{Condition} & \textbf{Composition Ratio} & \textbf{PSDS1$\uparrow$} & \textbf{PSDS2$\uparrow$} \\
        \midrule
        No Gen. (Baseline) & 6/6/0/6/12 & 0.4417 & 0.6639 \\
        \midrule
        +Gen. / -Real Strong & 3/6/3/6/12 & 0.4498 & 0.6861 \\
        +Gen. / -Synth. Strong & 6/3/3/6/12 & \underline{0.4606} & \underline{0.6877} \\
        +Gen. / -Real Weak & 6/6/3/3/12 & 0.4477 & 0.6817 \\
        +Gen. / -Unlabeled & 6/6/3/6/9 & 0.4519 & \textbf{0.6890} \\
        \midrule
        \rowcolor{lightblue} 
        +Gen. & 6/6/6/6/12 & \textbf{0.4641} & 0.6868 \\
        \bottomrule
    \end{tabular}
\end{table}

\section{Conclusion and Future Work}
In this work, we explore SynSonic, a data augmentation strategy for sound event detection that harnesses the diffusion ControlNet’s ability to generate single-event sound samples with precise control over temporal structures, enabling the synthesis of strongly labeled mixtures. To ensure the quality of the generated samples, we investigate a joint rank filtering strategy designed to mitigate the selection bias inherent in single-model filtering. Furthermore, we offer practical guidelines for effective integration of SynSonic, including mixing strategies and training batch composition. Experimental results show that this approach improves both PSDS1 and PSDS2 scores, reflecting gains in event localization and class discrimination. 

For future work, we plan to extend this approach to more complex sound event detection scenarios and explore the direct end-to-end generation of strongly labeled sound mixtures using generative models.

\clearpage
\bibliographystyle{IEEEtran}
\bibliography{refs25}







\end{document}